\documentclass[aps,amsmath,amssymb,showpacs,pre]{revtex4-1}
\usepackage{epsfig}
\usepackage{bm}
\usepackage{latexsym}
\usepackage{graphicx}
\usepackage{multirow}
\usepackage{diagbox}
\usepackage{amsmath}
\usepackage{caption}
\usepackage{color}
\usepackage{multirow}
\usepackage{setspace}
\usepackage{comment}

\usepackage{color}

\begin{document}
\title{Thermal Casimir interactions for higher derivative field Lagrangians:\\
generalized  Brazovskii models}
\author{David S. Dean$^{1,2}$, Bing Miao$^{3}$, Rudolf Podgornik$^{2,4,5,6}$}
\affiliation{(1) Univerit\'e Bordeaux, CNRS, LOMA, UMR 5798, F-33400 Talence, France}
\affiliation{(2) Kavli Institute for Theoretical Sciences, University of Chinese Academy of Sciences (UCAS), Beijing 100049, China}
\affiliation{(3) Center of Materials Science and Optoelectronics Engineering, College of Materials Science and Opto-Electronic Technology, University of Chinese Academy of Sciences (UCAS), Beijing 100049, China}
\affiliation{(4) School of Physical Sciences, University of Chinese Academy of Sciences (UCAS), Beijing 100049, China}
\affiliation{(5) CAS Key Laboratory of Soft Matter Physics, Institute of Physics,
Chinese Academy of Sciences (CAS), Beijing 100190, China}
\affiliation{(6)Department of physics, Faculty of mathematics and physics, University of Ljubljana, Jadranska 19, 1000 Ljubljana, Slovenia}

\bibliographystyle{apsrev}
\begin{abstract}
We examine the Casimir effect for free statistical field theories which have Hamiltonians with second order derivative terms. Examples of such Hamiltonians arise from models of non-local electrostatics, membranes with non-zero bending rigidities and field theories of the Brazovskii type that arise for polymer systems. The presence of a second derivative term means that new types of boundary conditions can be imposed, leading to  a richer phenomenology of interaction
phenomena. In addition zero modes can be generated that are not present in standard first derivative models, and it is these zero modes which give rise to long range Casimir forces. Two physically distinct cases are considered: (i) {\sl unconfined fields}, usually considered for finite size embedded inclusions in an infinite fluctuating medium, here in a two plate geometry the fluctuating field exists both inside and outside the plates, (ii) {\sl confined fields}, where the field is absent outside the slab confined between the two plates. We show how these two physically distinct cases are mathematically related and discuss a wide range of commonly applied boundary conditions. We concentrate our analysis to the critical region where the underlying bulk Hamiltonian has zero modes and show that very exotic Casimir forces can arise, characterised by very long range effects and oscillatory behavior that can lead to strong metastability in the system.
\end{abstract}
\maketitle

\section{Introduction}

Besides electrodynamic field fluctuations and the ensuing Casimir - van der Waals interactions, which can be captured by field theories with actions that contain at most first order (spatial and temporal) derivatives, many more complex systems have actions  which contain second order derivative terms  \cite{review2016}. In the quantum mechanics context higher derivative Lagrangians arise naturally when weak relativistic corrections to standard quantum mechanics are taken into account \cite{simon1990}. In  soft matter physics, Hamiltonians with second order derivative terms  are notably encountered in the context of stiff or semi-flexible polymers \cite{Bruinsma2013, pap, doi, dob2001, smith, uchida} that are constrained either in the embedding space or in the internal space coordinates due to the presence of adsorbing fluctuation quenchers. A natural context for higher derivative Hamiltonians is in soft (phospholipid) membranes  \cite{boal, dean2007, Bing2017}.  Liquid crystal theory also provides an interesting playground for exploring the fluctuation effects in the context of higher derivative Lagrangians \cite{had_2018, had_2019}. In addition, recent developments in the theory of ionic liquids, where finite ion sizes are important, lead to mean field theories which introduce higher order derivatives than the usual second order Poisson Boltzmann theory \cite{sant2006, ciach2007, bazant2011, blossey2017} and the analysis of the one loop correction to such theories will typically require higher order path integral formulations. In addition, effective field theories with higher derivative actions and even non-local field actions also arise from the dynamics of lower derivative theories in the dynamical context when analysing the evolution  of fluctuation forces toward their equilibrium values from the general non-equilibrium initial states \cite{ddaj1,ddaj2}. 

There have been several attempts to evaluate the field fluctuations or equivalently the field propagators in the case of higher derivative Lagrangians, most notably by Kleinert \cite{klein86} and his results are given in one of the standard textbooks on path integration \cite{handbook}. Some special cases of these results were established even before the developments of the full theory \cite{pap}. Recently the authors have established an alternative  derivation of Kleinert's results \cite{DMP-1} based on a link between unconfined systems, which can be treated using Green's function methods, and confined systems (which correspond to the standard path integral). The aim of this paper is to exploit these results to explore the Casimir interaction arising in both unconfined and confined geometries

For both confined and unconfined systems we examine the interaction between two parallel surfaces {for the Brazovskii model field Hamiltonian} \cite{Bra1,Bra2} with various boundary conditions imposed at each surface. For instance the obvious boundary conditions are {\sl Dirichlet} (D) and {\sl Neumann} (N), however for a second derivative action we can also apply D and N conditions at the same surface - the so called {\sl strong anchoring boundary condition}. We concentrate our study at the critical point where the model has a continuum of zero modes with wave vectors such that $|{\bf k}|=q_0$, {where $q_0$ is a parameter of the Brazovskii Hamiltonian}. At the critical point we see that, for both confined and unconfined systems, the modes in planes parallel to the surfaces make qualitatively different contributions to the Casimir interaction disjoining pressure. The modes with $q<q_0$, {where, $q=|{\bf q}|$}, can lead to oscillatory behavior, 
while giving the same contribution for confined and unconfined systems. Furthermore the Casimir interaction generated by these modes is of a 
longer range than usually seen in the thermal Casimir effect for free fields, giving a contribution to the disjoining pressure which decays with an envelope of $1/h$ ($h$ being the distance between the plates) rather than the usual $1/h^3$ behavior. The contribution to the Casimir interaction coming from the modes with $q>q_0$ leads to a standard thermal Casimir disjoining pressure, which decays as $1/h^3$ and is independent of the value of $q_0$.  Its amplitude and sign is a strong function of applied boundary conditions.

\section{Basic field theory}\label{basic}

We first describe the field theory in its bulk form which can be used to study unconfined systems and where the plates can be regarded as inclusions which impose constraints on the fluctuating order parameter. An approximation to the Ginzburg-Landau-Wilson Hamiltonian for {diblock} copolymer micro-phase separation is given by the Brazovskii model \cite{Fredrickson2006} where the Hamiltonian for the density-field fluctuations $\phi$ is given by
\begin{equation}
\beta{\cal H}[\phi] = \frac12 \int_V d{\bf r}\ [\nabla^2 \phi({\bf r})+q_0^2\phi({\bf r})]^2 + p_0^4 \phi({\bf r})^2,\label{form1}
\end{equation}
with $V$ the bulk volume of the system. We see that in the limit $p_0\to 0$ the system has zero-mode fluctuations corresponding to fluctuations with wave vectors ${\bf k}
$ such that $|{\bf k}|=q_0$. The modification of these zero and close to zero modes by the presence of boundaries should lead therefore to a strong Casimir effect. Another bulk Hamiltonian that can be naturally written down from a Landau-Ginsburg-Wilson perspective is
\begin{equation}
\beta{\cal H}'[\phi] = \frac12 \int_V d{\bf r}\ [\nabla^2 \phi({\bf r})]^2- 2q_0^2[\nabla\phi({\bf r})]^2 + (p_0^4+q_0^4) \phi({\bf r})^2,\label{form2}
\end{equation} 
which is clearly equivalent to the first Hamiltonian up to a surface term and trivially we see that
\begin{equation}
\beta{\cal H}'[\phi] = \beta{\cal H}[\phi]  - q_0^2\int_{\partial V}\phi({\bf r})\nabla\phi({\bf r})\cdot{\bf n} dS,
\end{equation}
where $\partial V$ denotes the surface of the system, ${\bf n}$ is the normal to the surface and $dS$ the area element. This second model with ${\cal H}'$ was the one studied by Uchida \cite{uchida}. The two models above  will clearly  have the same bulk behavior but interactions between embedded surfaces will in general be  modified by this surface term. However, in the case where either Dirichlet or Neumann boundary conditions are imposed at any of the surfaces and for periodic boundary conditions, the two models are clearly equivalent. In this paper we will carry out the analysis for the model given in Eq. (\ref{form1}), as the Hamiltonian is clearly positive for any field configurations, and can thus have zero modes but not strictly unstable ones. 

In what follows we will introduce planar boundaries perpendicular to the direction $z$. If we write ${\bf r}=(z,{\bf x})^T$ and then express the field in terms of Fourier modes in the subspace ${\bf x}$, we find the equivalent Hamiltonian in terms of real fields $\phi({\bf q},z)$ (using the fact  that the original field $\phi(z,{\bf x})$ is real).
\begin{equation}
\beta{\cal H}[\phi] = \sum_{\bf q} \frac{1}{2}\int_{z_1}^{z_2} dz \left[\ddot \phi(z,{\bf q})^2 + (\omega_1^2(q) + \omega_2^2(q) )
\dot \phi(z,{\bf q})^2  + \omega_1^2(q)\omega_2^2(q)\phi(z,{\bf q})^2\right] - \frac{(\omega_1^2(q)+\omega_2^2(q))}{2}\left[\dot \phi(z,{\bf q})\phi(z,{\bf q})\right]_{z_1}^{z_2}
\label{ragjm}
\end{equation}
where
\begin{equation}
\omega^2_1(q) = q^2-q_0^2 + i p_0^2\ {\rm and} \ \omega^2_2(q) = q^2-q_0^2 -i p_0^2,
\end{equation}
and $z_1$ denotes the left most point where the field is present an $z_2$ the right most point. We have written the Hamiltonian in a form familiar in polymer physics, the second derivate term being the polymer bending energy and the first derivative term being the stretching energy, while the last term corresponds to an overall confining quadratic potential. In the unconfined case we can take the limit $z_1\to-\infty$ and $z_2\to+\infty$. In this case one can set the surface term at $z=z_1$ and $z=z_2$ to zero, for instance by taking Dirichlet $\phi(z_i,{\bf q})=0$ or Neumann $\partial_z \phi(z_i,{\bf q}) = \dot\phi(z_i,{\bf q})=0$ boundary conditions. However, this does not affect the result as the plates are within the bulk and insensitive to this choice, provided that the correlation function of the field decays with distance.

A general Casimir set up with two plates, at $z=0$ and $z=h$, can be analysed by finding the solution to the problem where the fields $\phi(z,{\bf q})$ and $\dot \phi(z,{\bf q})$ are constrained  on both boundaries. This means that we need to compute the propagator
\begin{equation}
K(\phi,\dot \phi, \phi',\dot\phi',\omega_1,\omega_2,h) =
\int d[\phi]\delta(\phi(0)-\phi)\delta(\dot\phi(0)-\dot\phi)\delta(\phi(h)-\phi')\delta(\dot\phi(h)-\dot\phi')\exp(-\beta{ \cal H}_b(\omega_1,\omega_2))
\label{qfghj}
\end{equation}
where
\begin{equation}
\beta{ \cal H}_b(\omega_1,\omega_2)= \frac{1}{2}\int_{z_1}^{z_2} dz \left[\ddot \phi(z)^2 + (\omega_1^2 + \omega_2^2 )
\dot \phi(z)^2  + \omega_1^2\omega_2^2\phi(z)^2\right].
\label{fegjqykj}
\end{equation}
We will denote with $K_U(\phi,\dot \phi, \phi',\dot\phi',\omega_1,\omega_2,h)$ the propagator for the unconfined systems where $z_1\to-\infty $ and $z_2\to\infty $, while we denote by $K_C(\phi,\dot \phi, \phi',\dot\phi',\omega_1,\omega_2,h)$ the confined system where $z_1=0$ and $z_2=h$. {Note that the surface term in Eq. (\ref{ragjm}) then enters additively in the exponent of the propagator Eq. (\ref{qfghj}), computed solely with the bulk Hamiltonian Eq. (\ref{fegjqykj}).} 

\begin{figure}[t!]
\begin{center}
  \includegraphics[scale=0.35]{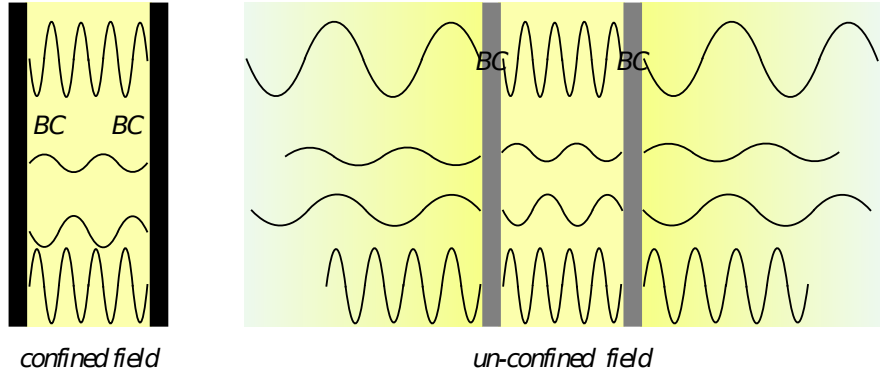} 
 \caption{Confinement of the fluctuations to different regions of the embedding space: confined vs. unconfined. In the confined case additional boundary conditions (BCs) can be imposed on the bounding surface(s), but not necessarily so. In the unconfined case the boundary conditions are essential and required. Without them the system is reduced to its bulk state.}\label{schematic}
\end{center}
\end{figure}

\begin{itemize}
\item In the {\bf confined case} Kleinert \cite{klein86} showed that 
{\begin{eqnarray}
K_C(\phi,\dot \phi, \phi',\dot\phi',\omega_1,\omega_2,h) &=& \frac{(\omega_1\omega_2)^\frac{1}{2}[(\omega_1^2-\omega_2^2)^2]^\frac{1}{2}}{2\pi M^{\frac{1}{2}}}\times \nonumber \\&&
\exp\left(-\frac{1}{2}\begin{pmatrix}&\phi'\\&\dot\phi'\end{pmatrix}\cdot S_D \begin{pmatrix}&\phi'\\&\dot\phi'\end{pmatrix} -\frac{1}{2}\begin{pmatrix}&\phi\\&-\dot\phi\end{pmatrix}\cdot S_D \begin{pmatrix}
&\phi\\&-\dot\phi\end{pmatrix} + \begin{pmatrix}&\phi\\&\dot\phi\end{pmatrix}\cdot S_C \begin{pmatrix}&\phi'\\&\dot\phi'\end{pmatrix}\right).\label{kleinert}
\end{eqnarray}}
Where here
\begin{equation}
M = (\omega_1^2+\omega_2^2) s_1 s_2 - 2\omega_1 \omega_2 c_1 c_2+2\omega_1\omega_2,
\end{equation}
\begin{equation}
S_D= \frac{1}{M}\left(
\begin{array}{cc}
 \omega_1\omega_2 (\omega_1^2-\omega_2^2) (\omega_1 c_2 s_1-\omega_2 c_1 s_2) &  \omega_1\omega_2
   \left( 2\omega_1 \omega_2s_1 s_2- \left(\omega_1^2+\omega_2^2\right) (c_1c_2-1)\right) \\ \omega_1\omega_2
   \left( 2\omega_1 \omega_2s_1 s_2- \left(\omega_1^2+\omega_2^2\right) (c_1c_2-1)\right)
   & (\omega_1^2-\omega_2^2)(\omega_1 c_1s_2-\omega_2 c_2  s_1) \\
\end{array}
\right)
\end{equation}
and
\begin{equation}
S_C= \frac{\omega_1^2 -\omega_2^2}{M}\begin{pmatrix}&\omega_1\omega_2(\omega_1 s_1 -\omega_2 s_2) & -\omega_1\omega_2(c_1-c_2) \\ & \omega_1\omega_2(c_1-c_2) & \omega_1 s_2 -\omega_2 s_1\end{pmatrix},
\end{equation}
and we have used the notation ${s_i=\sinh(\omega_ih)}$ and ${c_i=\cosh(\omega_ih)}$.
This result was rederived, in a very different way to \cite{klein86}, in \cite{DMP-1} by exploiting  a link with the unconfined case.

\item In the {\bf unconfined case} we showed in \cite{DMP-1} that
{\begin{eqnarray}
K_U(\phi,\dot \phi, \phi',\dot\phi',\omega_1,\omega_2,h) &=&  \frac{1}{\pi}(\omega_1\omega_2)^{\frac{1}{2}}(\omega_1+\omega_2)
\exp(\frac{1}{2} (\omega_1+\omega_2) h) K_C(\phi,\dot\phi,\phi',\dot\phi';h)\times \nonumber\\
&&
\exp\left(-\frac{1}{2}\begin{pmatrix}&\phi'\\&\dot\phi'\end{pmatrix}\cdot S_L \begin{pmatrix}&\phi'\\&\dot\phi'\end{pmatrix} -\frac{1}{2}\begin{pmatrix}&\phi\\&-\dot\phi\end{pmatrix}\cdot S_L \begin{pmatrix}
&\phi\\&-\dot\phi\end{pmatrix} \right)
\label{relation}
\end{eqnarray}}
where here
\begin{equation}
S_L= \begin{pmatrix} &\omega_1\omega_2(\omega_1+\omega_2) & \omega_1\omega_2 \\
& \omega_1\omega_2  & \omega_1 +\omega_2\end{pmatrix}.
\end{equation}
\end{itemize}
At this point it is interesting to compare the difference between the unconfined and confined path integrals. We see first of all that the  presence of the external bulk field with respect to the confined case leads to a difference in the overall prefactors independent of $h$, this difference is therefore unimportant for the Casimir effect. Secondly, with respect to  the unconfined case, the confined case has an additional factor $\exp(-\frac{1}{2} (\omega_1+\omega_2) h)$ which corresponds to an additional {\em bulk} free energy $\Delta F_{\bf q}= k_BT(\omega_1({q}) + \omega_2({q}))h/2$ per mode ${\bf q}$, implying a negative pressure per mode ${\bf q}$ and giving an additional bulk free energy
\begin{equation}
F_b = \frac{k_BThA}{2}\int \frac{d{\bf q}}{(2\pi)^2}(\omega_1({ q}) + \omega_2({q})),\label{bulk}
\end{equation}
where $A$  is the area of the plates. The corresponding pressure difference between the unconfined and confined cases is an excess bulk pressure due to the presence of the external bulk. We will see later that the bulk part of the pressure for the unconfined system actually turns out to be zero, which is normal as the bulk pressure of the interior is cancelled by that of the exterior as the volume occupied by the field is conserved upon changing the distance between the plates for the unconfined system.  Thirdly and most importantly, we see that the effective surface energy is renormalized due to the presence of additional quadratic terms in the vector $(\phi,\dot\phi,\phi',\dot\phi')^T$. Even though these terms do not depend explicitly on $h$, they are coupled to terms that do depend on $h$ and will thus lead to a modification of the Casimir pressure which we define as the $h$ dependent component of the pressure. 

This means that effective Hamiltonians for the problems where the fields are confined and unconfined respectively (but with the same boundary conditions on the two surfaces) can be written as 
\begin{equation}
\beta{\cal H}_C[\phi] = \sum_{\bf q} \frac{1}{2}\int_{0}^{h} dz \left[\ddot \phi(z,{\bf q})^2 + (\omega_1^2(q) + \omega_2^2(q) )
\dot \phi(z,{\bf q})^2  + \omega_1^2(q)\omega_2^2(q)\phi(z,{\bf q})^2\right] - \frac{(\omega_1^2(q)+\omega_2^2(q))}{2}\left[\dot \phi(z,{\bf q})\phi(z,{\bf q})\right]_{0}^{h}\label{hcon}
\end{equation}
while
\begin{eqnarray}
\beta{\cal H}_U[\phi] &=& \sum_{\bf q} \frac{1}{2}\int_{0}^{h} dz \left[\ddot \phi(z,{\bf q})^2 + (\omega_1^2(q) + \omega_2^2(q) )
\dot \phi(z,{\bf q})^2  + \omega_1^2(q)\omega_2^2(q)\phi(z,{\bf q})^2\right] \nonumber \\ &-&(\omega_1({q}) + \omega_2({q}))\frac{h}{2} + \omega_1(q)\omega_2(q)\left[\dot \phi(z,{\bf q})\phi(z,{\bf q})\right]_{0}^{h} +\frac{1}{2}(\phi(0,{\bf q})^2 + \phi(h,{\bf q})^2)\omega_1(q)\omega_2(q)(\omega_1(q) + \omega_2(q)) \nonumber \\&+&\frac{1}{2}(\dot\phi(0,{\bf q})^2 + \dot\phi(h,{\bf q})^2)(\omega_1(q) + \omega_2(q)) .\label{huc}
\end{eqnarray}
This is particularly interesting as the effect of external bulk can be represented in terms of a purely surface term in addition to the confined Hamiltonian along with a bulk free energy term, that is to say
\begin{eqnarray}
&&\beta \Delta{\cal  H}_B = \beta{\cal H}_U[\phi]- \beta{\cal H}_C[\phi] =
\nonumber\\
&&\sum_{\bf q} (\omega_1({q}) + \omega_2({q}))[-\frac{h}{2} + \frac{1}{2}(\omega_1(q)+\omega_2(q))\left[\dot \phi(z,{\bf q})\phi(z,{\bf q})\right]_{0}^{h} +\frac{1}{2}(\phi(0,{\bf q})^2 + \phi(h,{\bf q})^2)\omega_1(q)\omega_2(q) +\frac{1}{2}(\dot\phi(0,{\bf q})^2 + \dot\phi(h,{\bf q})^2)] \nonumber \\
.\label{hdiff}
\end{eqnarray}
However if one tries to write this equivalence in real space, the resulting surface energies (written as integrals over the surface) are non-local.

For the two cases, $q>q_0$ and $q<q_0$,  we have $\omega_{1}(q)=\sqrt{q^2\!\!-\!q_0^2+i p_0^2}$ and  $\omega_1(q)=i \sqrt{q_0^2-q^2-i p_0^2}$, respectively, with $\omega_2(q)=\omega_1^*(q)$,
in both cases . At the critical point defined as $p_0=0$, one finds that $\omega_1(q)=\omega_2(q)=\sqrt{q^2\!\!-\!q_0^2}$ are both real for $q>q_0$, while $\omega_1(q)=\omega_2^*(q)=i\sqrt{q_0^2-q^2}$ is purely imaginary for $q<q_0$. This has a  remarkable consequence in that for  $q<q_0$ we have $\omega_1(q)+\omega_2(q) =0$ and thus when $q<q_0$ the Hamiltonians ${\cal H}_U$ and ${\cal H}_C$ are identical!

We also see that  when $p_0=0$, the bulk free energy only contains contributions from the modes with $q>q_0$ and thus 
identifying in a standard fashion $\sum_\mathbf{q}=(A/(2\pi)^2)\int d\mathbf{q}$, we find 
\begin{equation}
F_b =  {k_BThA} \int_{q>q_0} \frac{d{\bf q}}{(2\pi)^2} \sqrt{q^2\!\!-\!q_0^2}\label{ebulk},
\end{equation}
which needs to be regularized by using an ultra-violet cut-off.

\section{Fluctuation induced interactions for confined fields}\label{confined}

We start with the confined field setup, where the field permeates only the slab  in the intersurface region and it is absent outside. The basic partition function for each mode, 
for fixed values { $\phi,\ \dot \phi,\ \phi',\ \dot \phi'$ }of the field and its normal derivatives on the surface, for this system can be written, using the form of the propagator in Eq. (\ref{kleinert}) and the expression (\ref{hcon}) as 
\begin{eqnarray}
K_C(\phi,\dot \phi, \phi',\dot\phi',\omega_1,\omega_2,h) &=& \frac{(\omega_1\omega_2)^\frac{1}{2}[(\omega_1^2-\omega_2^2)^2]^\frac{1}{2}}{2\pi M^{\frac{1}{2}}}\times \nonumber \\&&
\exp\left(-\frac{1}{2}\begin{pmatrix}&\phi'\\&\dot\phi'\end{pmatrix}\cdot S_{DR} \begin{pmatrix}&\phi'\\&\dot\phi'\end{pmatrix} -\frac{1}{2}\begin{pmatrix}&\phi\\&\dot\phi\end{pmatrix}\cdot P S_{DR} P\begin{pmatrix}
&\phi\\&\dot\phi\end{pmatrix} + \begin{pmatrix}&\phi\\&\dot\phi\end{pmatrix}\cdot S_C \begin{pmatrix}&\phi'\\&\dot\phi'\end{pmatrix}\right),\label{masterc}
\end{eqnarray}
where
\begin{equation}
S_{DR} = S_D - \frac{1}{2}\begin{pmatrix} &0& \omega_1^2 + \omega_2^2 \\
& \omega_1^2 + \omega_2^2 & 0 \end{pmatrix},
\end{equation}
incorporates the surface terms in Eq. (\ref{hcon}) and for notational convenience we have introduced the matrix
\begin{equation}
P = \begin{pmatrix}&1&0\\&0 & -1\end{pmatrix}.
\end{equation}
The corresponding free energy can be written in terms of the partition function $Z_C(\mathbf{q}, h)$ as
\begin{equation}
F(h)=- \frac{A k_BT}{(2\pi)^2} \int  d{\bf q}\ln\left(Z_C(\mathbf{q}, h)\right), 
\end{equation}
and incorporates the effects of the boundary conditions 
of the system as will become clear as we proceed. 

\subsection{Strong anchoring boundary condition}

The strong anchoring limit is defined as $\phi(z,{\bf x})=\dot\phi(z,{\bf x})\equiv \partial_z\phi(z,{\bf x})= 0$ at both surfaces. We also denote this by {\sl $DN-DN$ }boundary conditions.  The vanishing of the surface terms means that the partition function for the mode ${\bf q}$ is simply given by
\begin{equation}
{Z_C({\bf q}, h) = K_C(0,0,0,0,\omega_1,\omega_2,h) =  \frac{(\omega_1\omega_2)^\frac{1}{2}[(\omega_1^2-\omega_2^2)^2]^\frac{1}{2}}{2\pi M^{\frac{1}{2}}}}
\end{equation}
and from this we obtain at the critical point, where $p_0=0$,
\begin{eqnarray}
Z_C(\vert{\bf q}\vert >q_0, h) = \frac{q^2\!\!-\!q_0^2}{\pi (\sinh^2(t)-t^2)^{1/2}},
\end{eqnarray}
with $h(q^2\!\!-\!q_0^2)^{1/2}=t$. In the case where $q<q_0$ the partition function is 
\begin{eqnarray}
Z_C(\vert{\bf q}\vert<q_0, h) = \frac{q_0^2-q^2}{\pi (t^2-\sin^2 (t))^{1/2}},
\end{eqnarray}
where in this region  $t= h(q_0^2-q^2)^{1/2}$. The total free energy is then evaluated as 
\begin{equation}
F(h)=- \frac{A k_BT}{(2\pi)^2} \Big(\int_{q>q_0} \!\!\!\!\!d{\bf q}\ln\left(Z_C(\vert\mathbf{q}\vert>q_0, h)\right) + \int_{q<q_0} \!\!\!\!\!d{\bf q}\ln \left(Z_C(\vert\mathbf{q}\vert<q_0, h)\right)\Big)=F_>(h)+F_<(h), 
\end{equation}
where we have used $F_>$, $F_<$ to represent the free energy contributions from modes with $q>q_0$ and $q<q_0$, respectively. Using this we find that, up to terms independent of $h$, one has
\begin{equation}
F(h)=\frac{A k_BT}{2(2\pi)^2} \Big(\int_{q>q_0} \!\!\!\!\!d{\bf q}\ln(\sinh^2(t)-t^2)  + \int_{q<q_0} \!\!\!\!\!d{\bf q}\ln(t^2-\sin^2 (t))\Big) 
\end{equation}

Because in the confined case no field exists outside the slab between two bounding surfaces, the  free energy has a bulk term proportional to the volume coming from the divergent part in  term  $F_>(h)$, which we see from writing
\begin{equation}
F(h)=\frac{A k_BT}{2(2\pi)^2} \Big(\int_{q>q_0} \!\!\!\!\!d{\bf q}\ [\ln(4\exp(-2t)[\sinh^2(t)-t^2]) - \ln(4\exp(-2t))] + \int_{q<q_0} \!\!\!\!\!d{\bf q}\ln(t^2-\sin^2 (t))\Big) ,
\end{equation}
where the first integral above is now convergent, and from which one finds (again dropping $h$ independent terms), 
\begin{equation}
F(h)=\frac{A k_BT}{2(2\pi)^2} \Big(\int_{q>q_0} \!\!\!\!\!d{\bf q}\ln(4\exp(-2t)[\sinh^2(t)-t^2]) + \int_{q<q_0} \!\!\!\!\!d{\bf q}\ln(t^2-\sin^2 (t))\Big) + F_b
\end{equation}
where the bulk free energy is given in Eq. (\ref{ebulk}). One thus finds
\begin{equation}
F(h)-F_b=\frac{A k_BT}{4\pi h^2} \Big(\int_0^\infty  \!\!\!\!\!dt\ t\ln(4\exp(-2t)[\sinh^2(t)-t^2]) + \int_0^{q_0h} dt\  t\ln(t^2-\sin^2 (t))\Big).
\end{equation}
The  Casimir disjoining pressure  is then given by
\begin{equation}
\Pi = -\frac{\partial f(h)}{\partial h},
\end{equation}
where 
\begin{equation}
f(h) = \frac{F(h) - F_b}{A} = f_>(h) + f_<(h),
\end{equation}
and we find
\begin{equation}
f(h)=\frac{ k_BT}{2\pi h^2} \Big( -1.71629+ \frac{1}{2}\int_0^{q_0h} dt\  t\ln(t^2-\sin^2 (t))\Big).\label{safe}
\end{equation}
For $q>q_0$, the contribution to the disjoining pressure is
\begin{eqnarray}
\Pi_>(h) &=& -\frac{\partial f_>(h)}{\partial h} 
= -\frac{k_B T}{2\pi h^3} \int_0^\infty dt \frac{t^2[e^{-t}\sinh(t) + t^2 - t]}{\sinh^2 (t) - t^2},
\end{eqnarray}
while  for $q<q_0$ the contribution to the disjoining pressure is
\begin{eqnarray}
\Pi_<(h) &=&  -\frac{\partial f_<(h)}{\partial h} 
= -\frac{k_B T}{2\pi h^3} \int_0^{q_0 h} dt \frac{t^2 [t - \sin( t) \cos( t)]}{t^2 - \sin^2 (t)}.
\end{eqnarray}
The total pressure can  then be written as 
\begin{eqnarray}
\Pi(h) &=& \Pi_<+ \Pi_>  =  \frac{k_BT~q_0^3}{2\pi}  r(q_0 h),
\end{eqnarray}
{with
\begin{eqnarray}
r(x) &=& 
\frac{1}{x^3} \left( \int_0^{x}dt  \frac{t^2[\sin(t)\cos(t)-t]}{t^2-\sin^2(t)}-3.43258\right).
\label{fgejl}
\end{eqnarray}
This is the same as the result obtained from the Hamiltonian used by Uchida \cite{uchida} because, as pointed out above,  for these boundary conditions the two models used by Uchida and here are equivalent. The dimensionless Casimir disjoining pressure Eq. (\ref{fgejl}) shows very little structure, even if the first term is non-monotonic with
\begin{equation}
    -x^2 \leq \int_0^{x}dt  \frac{t^2[\sin(t)\cos(t)-t]}{t^2-\sin^2(t)} \leq -.5 x^2
\end{equation}
for the limits $x \ll 1$ and $x \gg 1$, displaying slight oscillations in between, however both the free energy and the pressure are monotonic and consequently the system exhibits no metastability.} 

\subsection{Robin Boundary Conditions}
Strong anchoring examined above corresponds to imposing  two boundary conditions at each surface. One can just as well impose a single boundary condition at each surface. For instance, imposing Robin boundary conditions on both surfaces means that for each mode, 
\begin{equation}
\left( b_i \dot\phi(z,{\bf q}) -\phi(z, {\bf q}) \right)\biggr\rvert_{z=0, z=h} = 0.
\end{equation}
In terms of the inward normal of the surfaces, these boundary conditions correspond to
\begin{equation} \phi(z,{\bf x})- b_1\nabla\phi(z,{\bf x})\cdot{\bf n}_1= 0\ \ {\rm and} \ \ 
\phi(z,{\bf x})+ b_2\nabla\phi(z,{\bf x})\cdot{\bf n}_2= 0.
\end{equation}
In this case the partition function for a single mode ${\bf q}$ is given by
\begin{equation}
Z_C(\mathbf{q}, h) = \int d\dot\phi~ d\dot\phi' ~K_C(b_1 \dot\phi, \dot\phi,b_2 \dot\phi',\dot\phi',\omega_1,\omega_2, h),
\end{equation}
and using {Eq. (\ref{masterc})} this yields
\begin{eqnarray}
&& Z_C(\mathbf{q}, h)= \frac{(\omega_1\omega_2)^\frac{1}{2}[(\omega_1^2-\omega_2^2)^2]^\frac{1}{2}}{2\pi M^{\frac{1}{2}}}\times \nonumber \\&&
\int d\dot\phi d\dot\phi'\exp\left(-\frac{\dot\phi'^2}{2}{\bf u}\cdot B_2S_{DR}B_2 {\bf u} -\frac{\dot\phi^2}{2}{\bf u}\cdot B_1P S_{DR}PB_1 {\bf u} + \dot\phi\dot\phi'{\bf u}\cdot B_2 S_C B_1{\bf u}\right),
\end{eqnarray}
where we have introduced a matrix and a vector as
\begin{equation}
B_i = \begin{pmatrix}&b_i&0\\&0 & 1\end{pmatrix} \ {\rm,} \ {\bf u}= \begin{pmatrix}&1\\&1\end{pmatrix}.
\end{equation}
Performing the integrations over $\dot\phi$ and $\dot\phi'$ we find
\begin{equation}
Z_C({\bf q}, h)= \frac{(\omega_1\omega_2)^\frac{1}{2}[(\omega_1^2-\omega_2^2)^2]^\frac{1}{2}}{M^{\frac{1}{2}}}
\left[ \left( {\bf u}\cdot B_2S_{DR}B_2 {\bf u} \right)\left({\bf u}\cdot B_1P S_{DR}PB_1 {\bf u} \right)-\left({\bf u}\cdot B_2 S_C B_1{\bf u}\right)^2\right]^{-\frac{1}{2}}\label{genrobin}
\end{equation}
The parameters $b_i$ introduce additional length scales and the general expression Eq. (\ref{genrobin}) is thus extremely complicated for general $b_1$ and $b_2$, below we thus restrict ourselves to a restricted set of parameters. 

We now first consider the symmetric case where $b_1=b_2=b$. Here we obtain the remarkably simple result
\begin{equation}
Z_C({\bf q}, h)= \left(\frac{\omega_1 \omega_2}{\sinh(\omega_1h)\sinh(\omega_2 h)(b^2\omega_1^2 -1)(b^2\omega_2^2 -1)}\right)^{\frac{1}{2}}.
\end{equation}
One can verify that the argument in the square root is positive definite due to the fact that $\omega_2=\omega_1^*$. We thus see that the $h$ dependent part of the free energy is independent of {$b$}, therefore for 
the whole range of $b$, from $b=0$ (D-D boundary conditions) through to $b=\pm\infty$ (N-N boundary conditions) the {\sl Casimir pressure is universal}. The effect of $b$ only appears in a term that corresponds to a surface energy. The above factorization of the  partition function can be understood in terms of a Fourier expansion of the second order path integral which can be shown to factorize into two first order integrals with the same boundary conditions. This holds because there is only one boundary condition at each surface. 

At the critical point we find that the Casimir interaction free energy (the $h$ dependent part) is given by
\begin{equation}
F(h) = \frac{Ak_BT}{2 (2\pi)^2} \Big( \int_{q>q_0} \!\!\!d{\bf q}~ \ln{\left(\sinh^2{\left(h\sqrt{q^2\!\!-\!q_0^2}\right)}\right)} + \int_{q<q_0} \!\!\!d{\bf q}~ \ln{\left(\sin^2{\left(h\sqrt{q_0^2-q^2}\right)}\right)} \Big).
\label{bcwk}
\end{equation}
Subtracting the bulk free energy and again introducing the variable $t=h\sqrt{q_0^2-q^2}$ for $q<q_0$ and  $t=h\sqrt{q^2\!\!-\!q_0^2}$ for $q>q_0$, we obtain
\begin{eqnarray}
f(h)&=& \frac{k_BT}{2\pi h^2} \left( \int_{0}^{\infty} \!\!\! td{t}~ \ln{\left( 1 - \exp(-2t) \right)} + \int_{0}^{h q_0} \!\!\!td{t}~ \ln{\vert\sin{t}\vert} \right) \nonumber \\
&=& \frac{k_BT}{2\pi h^2} \left( -\frac{\zeta(3)}{4} + \int_{0}^{h q_0} \!\!\!d{t}\ t \ln{\vert\sin{t}\vert} \right),
\label{befjkw0}
\end{eqnarray}
where $\zeta(s)$ is the Riemann zeta function ($\zeta(3)=1.202$). This result agrees with that of Uchida \cite{uchida} who gave the case of Dirichlet boundary conditions only. The first term is simply twice the free energy, per unit area, due to the universal thermal (attractive) Casimir effect for massless scalar fields with Dirichlet-Dirichlet boundary conditions, and is, interestingly, independent of $q_0$. The second term is oscillatory and long-range with respect to the usual thermal Casimir effect.
\begin{figure}[h!]
	\begin{center}
		\includegraphics[width=12cm]{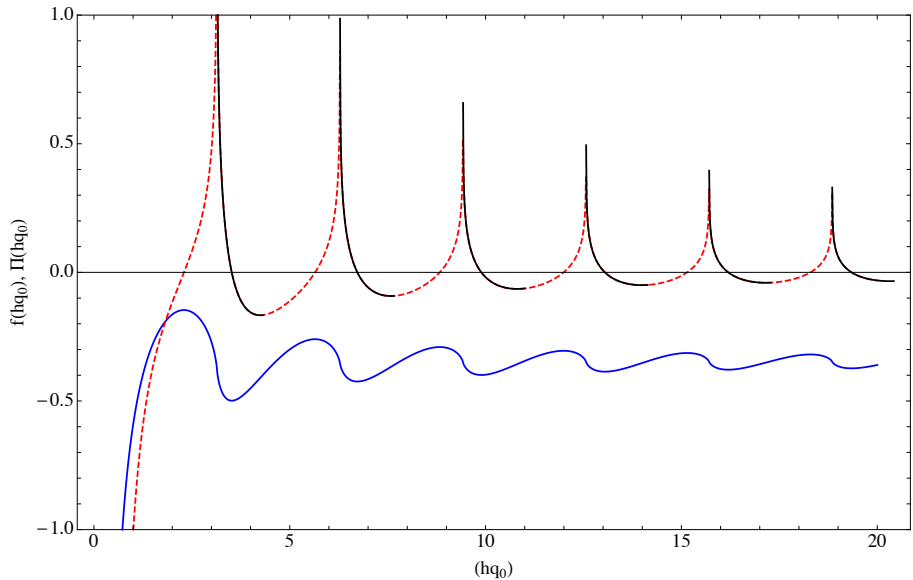}
		\caption{{Dimensionless Casimir free energy at the critical point, {$f(h) \rightarrow f(h)/\frac{k_BT~q_0^2}{(2\pi)}$} (blue - lower curve), from Eq. (\ref{befjkw0}) and dimensionless Casimir disjoining pressure $\Pi(h) \rightarrow \Pi(h)/\frac{k_BT~q_0^3}{(2\pi)}$ (red - upper curve) from Eq. (\ref{fahtw}) for the symmetric boundary conditions,} both as functions of dimensionless separation $h q_0$. The stable branches of disjoining pressure are indicated by the solid curve, and the unstable by dashed curve. The thermodynamic state is at $h=0$, but there also exist an infinite number of metastable states in the stable regions, where $\partial\Pi(h)/\partial h <0$ (black - solid curve), bounded by spikes of infinite pressure. The envelope of the free energy indicates an underlying attractive Casimir interaction.} 
		\label{fig:fig3pdd}
	\end{center}
\end{figure}

The absolute value inside the {logarithm} function is a consequence of the squared term in Eq. (\ref{bcwk}). The second integral in the bracket is related to the two famous Euler integrals as well as the Clausen function. The interaction free energy is finite for any non-zero separation, has a monotonic (attractive) envelope, but contains an infinite sequence of asymmetric local minima and maxima, separated by an infinite derivative. In the constant pressure ensemble, the position of thermodynamic equilibrium will depend on the total applied bulk pressure $P_t$ of the system and $h$ is determined from the solution of  the equation $\Pi(h) = P_t$. 

{Apart from the regular part scaling as $h^{-3}$ and $h^{-1}$ \cite{Ziherl1,Ziherl2}, in the vicinity of $hq_0 = n \pi$, $n = 1, 2, 3 \dots$,  the fluctuation-induced interaction pressure becomes repulsive and diverges logarithmically,
\begin{eqnarray}
\Pi(h) &=& - \frac{\partial f(h)}{\partial h} = 
\frac{2 f(h)}{h} - \frac{k_B T ~q_0^2}{2\pi h} \ln{\vert \sin{h q_0}\vert}.  
~
\label{fahtw}
\end{eqnarray}
As a result of this the  equation $\Pi(h) = P_t$ will have an infinite number of solutions for $P_t$ positive (a net applied pressure acting inward on the system) and even for a range of negative pressures as can be see from Fig \ref{fig:fig3pdd}. The thermodynamic stable state is, however, given  by $h=0$, while there also exists an infinite number of metastable states 
in the stable regions where $\partial\Pi(h)/\partial h <0$ as shown in Fig. \ref{fig:fig3pdd}.}


In the limit where $b_1=b\to \infty$ while $b_2=0$ we  obtain the case of Dirichlet-Neumann 
boundary conditions. Explicitly we have 

\begin{eqnarray}
Z_C(\mathbf{q}, h) =   \left(b^2 \cosh(h\omega_1) \cosh(h\omega_2)\right)^{-\frac{1}{2}},
\end{eqnarray}
and this factorisation can again be understood in terms of an eigenfunction expansion.

\begin{figure}[h!]
	\begin{center}
		\includegraphics[width=12cm]{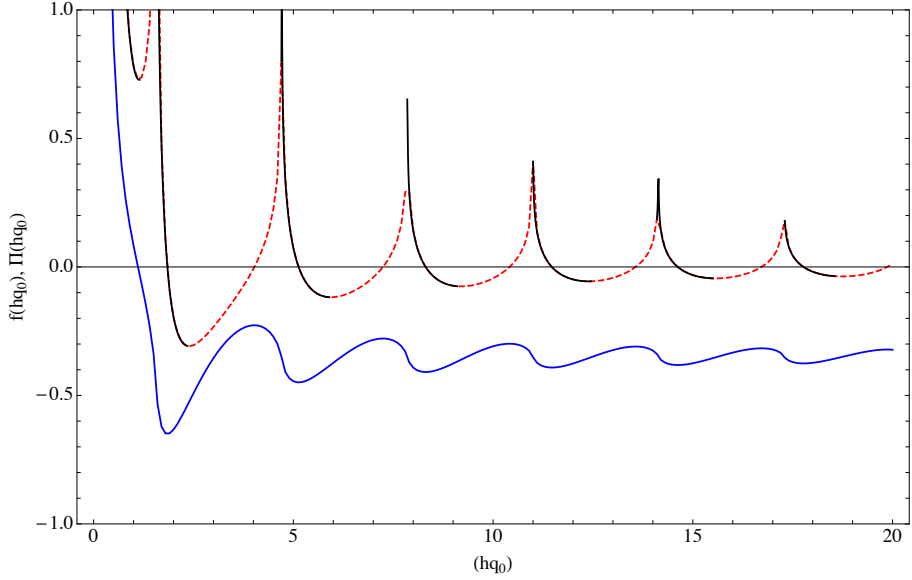}
		\caption{Dimensionless free energy at the critical point, $f(h) \rightarrow f(h)/\frac{k_BT~q_0^2}{(2\pi)}$ (blue - lower curve), from Eq. (\ref{dnc}) for Dirichlet-Neumman boundary conditions and the corresponding dimensionless disjoining pressure $\Pi(h) \rightarrow \Pi(h)/\frac{k_BT~q_0^3}{(2\pi)}$ (red - upper curve) from Eq. (\ref{mituyln}), both as functions of dimensionless separation $h q_0$. The stable branches of disjoining pressure are again indicated by the solid curve, and the unstable by dashed curve. The thermodynamic stable state is at a finite $h$, but there exits an infinite number of metastable states in the stable regions, where $\partial\Pi(h)/\partial h <0$ (black - solid curve), delimited by  infinite pressure spikes. The envelope of the free energy indicates an underlying repulsive Casimir interaction.}
		\label{fig:fig3p1}
	\end{center}
\end{figure}

At the critical point we find that the $h$ dependent part of the free energy is given by
\begin{equation}
F(h) =  \frac{Ak_BT}{2 (2\pi)^2} \Big( \int_{q>q_0} \!\!\!d{\bf q}~ \ln{\left(\cosh^2\left({h\sqrt{q^2\!\!-\!q_0^2}}\right)\right)} + \int_{q<q_0} \!\!\!d{\bf q}~ \ln{\left(\cos^2\left({h\sqrt{q_0^2-q^2}}\right)\right)} \Big),
\end{equation}
and this leads to 
\begin{eqnarray}
f(h) &=& \frac{k_BT}{2\pi h^2} \left( \int_{0}^{\infty} \!\!\! td{t}~ \ln{\left( 1 + \exp(-2t) \right)} + \int_{0}^{h q_0} \!\!\!d{t}\ t \ln{\vert\cos{t} \vert} \right) 
\nonumber\\
&=& \frac{k_BT}{2\pi h^2} \left( \frac{3\zeta(3)}{16} + \int_{0}^{h q_0} \!\!\!d{t}\ t\ln{\vert\cos{t} \vert} \right).
\label{dnc}
\end{eqnarray}
From this we see that
\begin{equation}
\Pi(h) = \frac{2f(h)}{ h} - \frac{k_BTq_0^2}{2\pi h}\ln\left({\vert\cos{q_0h} \vert} \right).
\label{mituyln}
\end{equation}
In contrast to the symmetric case given in Eq. (\ref{befjkw0}), we see that the monotonic contribution from the modes $q>q_0$ is repulsive (and has the form of twice the 
Dirichlet-Neumann Casimir force for massless first order Hamiltonians). The minimum free energy $f(h)$ occurs at a finite separation $h$, which means one can achieve the stable state at a finite slab depth. The equation $\Pi(h) = P_t$ in this case has no solution for sufficiently negative $P_t$, as can be seen from Fig. \ref{fig:fig3p1}, but again displays an infinite number of metastable states for $P_t$ positive with $\partial\Pi(h)/\partial h <0$. The oscillatory part of the pressure diverges at $hq_0 = (2n+1) {\textstyle\frac{\pi}{2}}$, $n = 0,1,2 \dots$, and leads to a quite exotic behavior as can be seen in Fig. \ref{fig:fig3p1}. The  envelope of the free energy is monotonic and overall repulsive. A behavior reminsicent of this, but without a diverging pressure, is found in the one dimensional Coulomb gas with charge regulation \cite{ddc}.

\subsection{Some more exotic boundary conditions}
\label{Secexpot}

Up to now we have considered the cases of strong anchoring, where all fields are constrained  to zero, meaning two constraints at each surface, and Robin boundary conditions, where a single constraint is imposed on each surface. Now we consider the {case} of strong anchoring on one surface (surface 2) and Robin boundary conditions on the other, written as  $\phi(0,{\bf x}) =b_1{\bf n}\cdot\nabla\phi |_{z=0}$. Here we find 
\begin{eqnarray}
Z_C({\bf q},h) &=& \frac{(\omega_1\omega_2)^\frac{1}{2}[(\omega_1^2-\omega_2^2)^2]^\frac{1}{2}}{2\pi M^{\frac{1}{2}}} \int d\dot\phi\exp\left(-\frac{\dot\phi^2}{2}{\bf u}\cdot  B_1P S_{DR}P B_1 {\bf u}\right)\nonumber \\
&=& \frac{(\omega_1\omega_2)^\frac{1}{2}[(\omega_1^2-\omega_2^2)^2]^\frac{1}{2}}{(2\pi)^{\frac{1}{2}} M^{\frac{1}{2}} ({\bf u}\cdot B_1 P S_{DR}P B_1{\bf u})^{\frac{1}{2}} }
\end{eqnarray}
The case for arbitrary $b_1$ is rather complicated due to the extra length scale introduced, however for Dirichlet boundary conditions $b_1=0$ (DN-D) we find
\begin{equation}
Z_C({\bf q},h)= \left(\frac{ \omega_1\omega_2 (\omega_1^2-\omega^2_2)}{
2\pi[\omega_1\cosh(\omega_1 h)\sinh(\omega_2 h) - \omega_2\cosh(\omega_2 h)\sinh(\omega_1 h)]}\right)^{\frac{1}{2}}
\end{equation}
At the critical point the interaction free energy, {\sl i.e.}, the $h$ dependent part, is given by 
\begin{equation}
F(h) = \frac{Ak_BT}{2 (2\pi)^2} \Big( \int_{q>q_0} \!\!\!d{\bf q}~ \ln\left(\sinh(2\sqrt{q^2-q_0^2}h)-2\sqrt{q^2-q_0^2}h\right) + \int_{q<q_0} \!\!\!d{\bf q}~ \ln\left(2\sqrt{q_0^2-q^2}h -\sin(2\sqrt{q_0^2-q^2}h)\right) \Big).
\end{equation}

\begin{figure}[t!]
	\begin{center}
		\includegraphics[width=12cm]{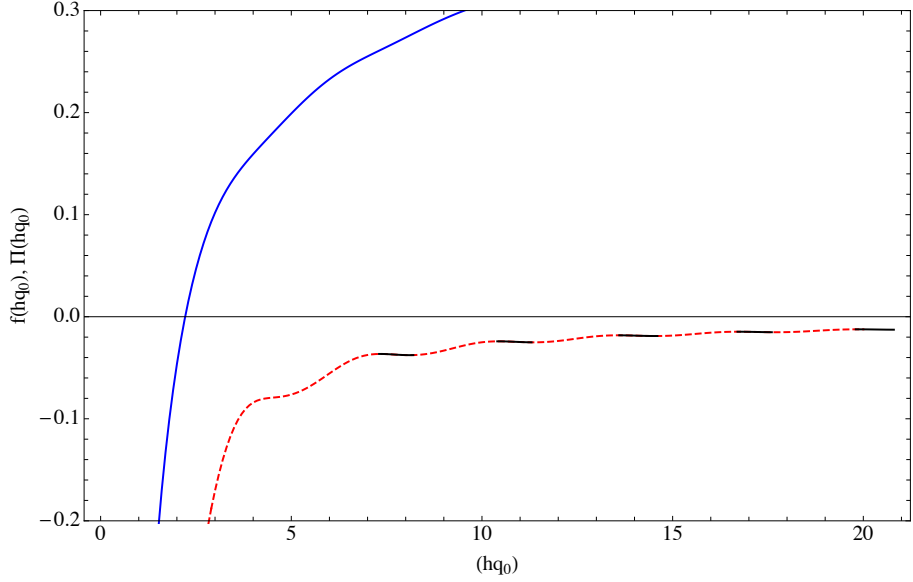}
		\caption{{Dimensionless free energy at the critical point for strong anchoring on one surface and Dirichlet boundary condition on the other, $f(h) \rightarrow f(h)/\frac{k_BT~q_0^2}{ 2(2\pi)}$, 
		(blue - upper curve) from Eq. (\ref{gdxhjkj}), and the corresponding dimensionless disjoining pressure $\Pi(h) \rightarrow \Pi(h)/\frac{k_BT~q_0^3}{(2\pi)}$ (red - lower curve) from Eq. (\ref{vadfghwjgg}), both as functions of dimensionless separation $h q_0$. The stable branches of disjoining pressure are again indicated by the solid curve, and the unstable by dashed curve. The thermodynamic stable state is at $h=0$, but there exits also an infinite number of metastable states in the stable regions, where $\partial\Pi(h)/\partial h <0$ (black - solid curve), accessible for negative $P_t$. The envelope of the free energy indicates an underlying attractive Casimir interaction.}}
		\label{fig:fig4p1}
	\end{center}
\end{figure}

Subtracting the bulk free energy then yields 
\begin{eqnarray}
f(h) &=& \frac{k_BT}{2\pi h^2}\left[\frac{1}{2}\int_0^\infty dt\  t \ln\left(1-\exp(-4t)-4 t\exp(-2t)\right)
+\frac{1}{2}\int_0^{q_0h}dt\ t \ln\left(2t-\sin(2t)\right)\right] \nonumber \\
&=&\frac{k_BT}{2\pi h^2}\left[-0.83591
+\frac{1}{2}\int_0^{q_0h}dt\  t \ln\left(2t-\sin(2t)\right)\right].
\label{gdxhjkj}
\end{eqnarray}
As in the case of strong anchoring, we find that the modes with $q>q_0$ give a contribution to the free energy which is attractive with a non-universal amplitude. 
The disjoining pressure is then given by
\begin{equation}
\Pi(h)= \frac{2f(h)}{h} -\frac{k_BT q_0^2}{4\pi h}\ln\left(2hq_0-\sin(2h q_0)\right).
\label{vadfghwjgg}
\end{equation}
{We observe, see Fig. \ref{fig:fig4p1}, that while the free energy is overwhelmingly determined by the monotonically attractive first term in Eq. (\ref{gdxhjkj}), the pressure, Eq. (\ref{vadfghwjgg}), shows that most values of $h$ are unstable with $\partial \Pi(h)/\partial h >0$ and the absolute minimum value of  $f(h)$ is at $h=0$. Numerical evaluation of $\partial \Pi(h)/\partial h $ does however reveal narrow regions of stability where $\partial \Pi(h)/\partial h <0$ which can thus be metastable at not too large negative applied pressures. The equation $\Pi(h) = P_t$ has a solution for any negative $P_t$ as can be seen from Fig. \ref{fig:fig4p1}, and in addition displays an infinite number of periodic van der Waals-like loops, if one applies the Maxwell rule. The oscillatory part of the pressure shows no divergence but nevertheless leads to an unusual behavior as can be seen in Fig. \ref{fig:fig4p1}. The  envelope of the free energy is monotonic and overall attractive.}

For Neumann boundary conditions, $b_1\to\infty$,  we find 
\begin{equation}
Z_C({\bf q},h)= \left(\frac{\omega_1^2-\omega_2^2}{
2\pi b_1^2[\omega_1\sinh(\omega_1 h)\cosh(\omega_2 h) - \omega_2\sinh(\omega_2 h)\cosh(\omega_1 h)]}\right)^{\frac{1}{2}}
\end{equation}
and here
\begin{equation}
F(h) = \frac{Ak_BT}{2 (2\pi)^2} \Big( \int_{q>q_0} \!\!\!d{\bf q}~ \ln\left(\sinh(2\sqrt{q^2-q_0^2}h)+2\sqrt{q^2-q_0^2}h\right) + \int_{q<q_0} \!\!\!d{\bf q}~ \ln\left(2\sqrt{q_0^2-q^2}h +\sin(2\sqrt{q_0^2-q^2}h)\right) \Big).
\end{equation}
Upon subtracting the bulk energy we find
\begin{eqnarray}
f(h) &=& \frac{k_BT}{2\pi h^2}\left[\frac{1}{2}\int_0^\infty dt ~t \ln\left(1-\exp(-4t)+4 t\exp(-2t)\right)
+\frac{1}{2}\int_0^{q_0h}dt\ t \ln\left(2t+\sin(2t)\right)\right]\nonumber \\
&=& \frac{k_BT}{2\pi h^2}\left[0.406839
+\frac{1}{2}\int_0^{q_0h}dt\  t \ln\left(2t+\sin(2t)\right)\right]
\label{bgdhj678}
\end{eqnarray}
Here we see that the modes with $q>q_0$ give a repulsive Casimir free energy,  with the standard $1/h^2$ behaviour but again with a non universal amplitude. The disjoining pressure in this case is given in complete analogy with the previous cases by
\begin{equation}
\Pi(h)= \frac{2f(h)}{h} -\frac{k_BT q_0^2}{4\pi h}\ln\left(2hq_0+\sin(2h q_0)\right)
\label{bgdhj679}
\end{equation}

\begin{figure}[t!]
	\begin{center}
		\includegraphics[width=12cm]{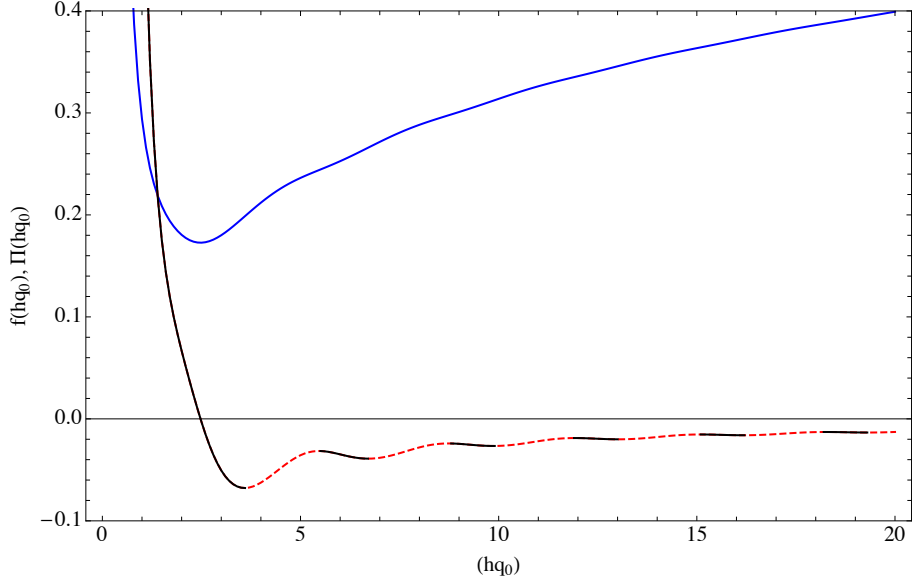}
		\caption{Dimensionless free energy at the critical point, $f(h) \rightarrow f(h)/\frac{k_BT~q_0^2}{2(2\pi)}$, 
		(blue-upper curve), from Eq. (\ref{bgdhj678}) for strong anchoring on one surface and Neumann boundary condition on the other and the corresponding dimensionless disjoining pressure $\Pi(h) \rightarrow \Pi(h)/\frac{k_BT~q_0^3}{(2\pi)}$ (red - dashed) from Eq. (\ref{bgdhj679}), both as functions of dimensionless separation $h q_0$. The stable branches of disjoining pressure are again indicated by the solid curve, and the unstable by dashed curve. The thermodynamic {stable} state is now at a finite value of $h q_0$. The equation $\Pi(h) = P_t$ has a single solution for any positive $P_t$, but can have multiple solutions for $P_t$ small and negative, corresponding to an infinite number of metastable states in the stable regions, where $\partial\Pi(h)/\partial h <0$ (black - solid curve). The envelope of the free energy indicates a non-monotonic Casimir interaction.}
		\label{fig:fig5p1}
	\end{center}
\end{figure}
In this case we observe, see Fig. \ref{fig:fig5p1}, that the free energy is dominated by the monotonically repulsive Casimir interaction stemming from the first term in Eq. (\ref{bgdhj678}) for small $h$, while for larger $h$ the second term takes over, leading to attractive Casimir interactions scaling approximately as $h^{2.15}$. The global minimum is achieved at a finite value of the separation $h$. The pressure, Eq. (\ref{bgdhj679}), exhibits an oscillatory component superimposed on a non-monotonic background and again shows an infinite sequence of regions in the metastable regime where $\partial\Pi(h)/\partial h <0$ separated by  unstable regions. The equation $\Pi(h) = P_t$ has a single solution for any positive $P_t$, while it can have multiple solutions for $P_t$ small and negative. The overall envelope of the free energy indicates a non-monotonic Casimir interaction.

Another possibility is that no boundary conditions are applied on a given surface. For example one can take strong anchoring boundary conditions on one surface and use free boundary conditions at the other. We note that for an unconfined system this choice of boundaries cannot lead to any interaction, as it corresponds to a single interface in the system. However for
confined systems there can still be an interaction. In this case we actually find
\begin{equation}
Z_C({\bf q},h)=\frac{(\omega_1\omega_2)^\frac{1}{2}[(\omega_1^2-\omega_2^2)^2]^{\frac{1}{2}}}{ M^\frac{1}{2}[{\rm det}(S_{DR})]^{\frac{1}{2}}} = \left(1-\frac{M}{4 \omega_1\omega_2}\right)^{-\frac{1}{2}}
\end{equation}
and in general we see that $Z_C({\bf q},h)$ will depend on $h$ through $M$. However at the critical point $p_0=0$ we find that for all $q$, $Z_C({\bf q},h)=1$ and so there is {\sl no interaction} between a surface with strong anchoring boundary conditions and another with completely free boundary conditions! In general one can show that at the critical point, if one surface has free boundary conditions, there is no interaction between the surfaces. 

For confined systems, periodic and antiperiodic boundary conditions are also relevant. Indeed the most general periodic/antiperiodic boundary conditions of this type can be written as
\begin{equation}
\begin{pmatrix}&\phi(h,{\bf x})\\ &\dot\phi(h,{\bf x})\end{pmatrix} =R\begin{pmatrix}&\phi(0,{\bf x})\\& \dot\phi(0,{\bf x})\end{pmatrix},
\end{equation}
where the matrix $R$ can take $4$ distinct forms
\begin{equation}
R(\sigma,\sigma') = \begin{pmatrix}&\sigma  & 0\\ & 0 & \sigma' \end{pmatrix},
\end{equation}
where $\sigma$ and $\sigma'=\pm 1$. The partition function can then be derived as 
\begin{eqnarray}
Z_C({\bf q},h) &=& \frac{(\omega_1\omega_2)^\frac{1}{2}[(\omega_1^2-\omega_2^2)^2]^\frac{1}{2}}{2\pi M^{\frac{1}{2}}}\times \nonumber 
\int d\phi~ d\dot\phi~\exp\left(-\frac{1}{2}\begin{pmatrix}&\phi\\&\dot\phi\end{pmatrix}\cdot[ RS_{DR}R+
PS_{DR}P-2S_CR] \begin{pmatrix}&\phi\\&\dot\phi\end{pmatrix} \right)\nonumber \\
&=& \frac{(\omega_1\omega_2)^\frac{1}{2}[(\omega_1^2-\omega_2^2)^2]^\frac{1}{2}}{ M^{\frac{1}{2}}{\rm det}\left( RS_{DR}R+PS_{DR}P - S_CR-R S_C^T\right)^\frac{1}{2}},
\end{eqnarray}
and while carrying out the Gaussian integration we must use the symmetric part of the relevant matrix. The corresponding partition functions are obtained with an obvious notation as
\begin{eqnarray}
Z_{C}({\bf q},h,++)&=& \frac{1}{4\sinh(\frac{h \omega_1}{2})\sinh(\frac{h \omega_2}{2})}\\
Z_{C}({\bf q},h,--)&=& \frac{1}{4\cosh(\frac{h \omega_1}{2})\cosh(\frac{h \omega_2}{2})}\\
Z_{C}({\bf q},h,+-)&=& Z_{C}({\bf q},-+)=\left[ -\frac{\omega_1 \omega_2}{M}\right]^\frac{1}{2},
\end{eqnarray}
wherefrom we can derive the interaction free energies as
\begin{eqnarray}
f(h,++)  &=& \frac{ k_BT}{2\pi h^2} \left(-2\zeta(3)+ 8\int_{0}^{h q_0/2} \!\!\!dt\ t \ln{\vert \sin{t}\vert} \right)\\
f(h,--)&=& \frac{ k_BT}{2\pi h^2} \left( \frac{3\zeta(3)}{2} + 8\int_{0}^{h q_0/2} \!\!\!dt\ t \ln{\vert\cos{t}\vert} \right), 
\end{eqnarray}
{and furthermore we see that $f(h,+-)$ is identical to the strong anchoring result given in Eq. (\ref{safe}). Furthermore, the periodic case is identical to twice the symmetric case Eq. (\ref{befjkw0}) and the antiperiodic case to twice the non symmetric case Eq. (\ref{dnc}), but evaluated at  $h/2$ as opposed to $h$. We shall thus not dwell on the details of these results.}

\section{Unconfined fields}

We now consider systems where the field exists both between and outside the plates. As we have  seen in section (\ref{basic}) the confined and unconfined systems differ by surface terms and in all but the strong anchoring case they will be different.

Here the partition function for each mode, 
for fixed values $\phi,\ \dot \phi,\ \phi',\ \dot\phi'$ of the field and its normal derivatives on the surface, is given using Eq. (\ref{kleinert}) and {(\ref{relation})} as 
\begin{eqnarray}
Z_U(\phi,\dot\phi,\phi',\dot\phi';\omega_1,\omega_2,h) &=& \frac{\omega_1\omega_2(\omega_1+\omega_2)([(\omega_1^2-\omega_2^2)^2]^\frac{1}{2}\exp(\frac{1}{2} (\omega_1+\omega_2) h)}{2\pi^2 M^{\frac{1}{2}}}\times \nonumber \\&&
\exp\left(-\frac{1}{2}\begin{pmatrix}&\phi'\\&\dot\phi'\end{pmatrix}\cdot S_{DR} \begin{pmatrix}&\phi'\\&\dot\phi'\end{pmatrix} -\frac{1}{2}\begin{pmatrix}&\phi\\&\dot\phi\end{pmatrix}\cdot P S_{DR} P\begin{pmatrix}
&\phi\\&\dot\phi\end{pmatrix} + \begin{pmatrix}&\phi\\&\dot\phi\end{pmatrix}\cdot S_C \begin{pmatrix}&\phi'\\&\dot\phi'\end{pmatrix}\right),\label{masterc1}
\end{eqnarray}
where here 
\begin{equation}
S_{DR} = S_D + S_L.
\end{equation}
Using this, one can read off the results from the various cases investigated for the confined field problem. 

The disjoining Casimir pressure for the strong anchoring case is identical to that for confined systems as the only difference is the bulk term and $h$ independent terms, which can be interpreted as surface free energies. In what follows there is no need to subtract the bulk pressure as it is automatically subtracted due to the fact that the field exists both outside and between the plates. Below we consider the same set  of boundary conditions as considered above for confined fields. As the basic idea of the calculations are given in Section (\ref{confined}) , we do not give the full details in what follows.
\begin{figure}[t!]
	\begin{center}
		\includegraphics[width=12cm]{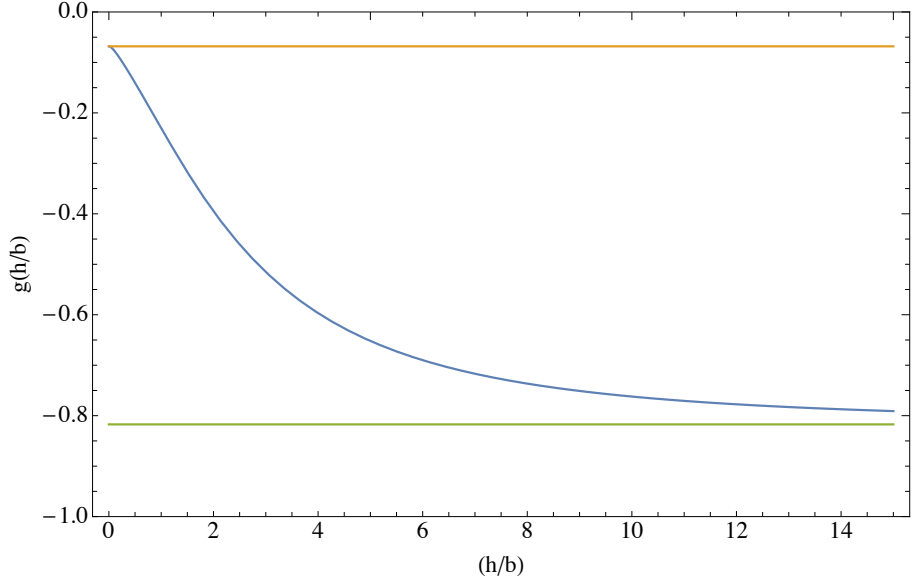}
		\caption{{The dependence of the first integral in Eq. (\ref{bakxruyb}) on the ratio $h/b$. The two asymptotes show the value of the integral for D-D boundary conditions ($b=0$) and N-N boundary conditions ($b=\infty$), respectively, see Eqs. (\ref{mwegh}) and (\ref{bfgjq}).}}
		\label{fig:fig6p1}
	\end{center}
\end{figure}

\subsection{Robin boundary conditions}

Using the same notation as Section (\ref{confined}) we find
\begin{equation}
Z_U({\bf q}, h)= \frac{\omega_1\omega_2(\omega_1+\omega_2)([(\omega_1^2-\omega_2^2)^2]^\frac{1}{2}\exp(\frac{1}{2} (\omega_1+\omega_2) h)}{\pi M^{\frac{1}{2}}}
\left[ \left( {\bf u}\cdot B_2S_{DR}B_2 {\bf u} \right)\left({\bf u}\cdot B_1P S_{DR}PB_1 {\bf u} \right)-\left({\bf u}\cdot B_2 S_C B_1{\bf u}\right)^2\right]^{-\frac{1}{2}},\nonumber \\ \label{genrobinu}
\end{equation}
which for symmetric Robin boundary conditions, where $b_1=b_2=b$, yields
\begin{equation}
f(h)= \frac{k_B T }{2(2\pi)^2}\int d{\bf q}\ln\left(1 - \frac{e^{-2h(\omega_1+\omega_2)}\left(e^{h\omega_2}\omega_2(b^2\omega_1^2-1)-e^{h\omega_1}\omega_1(b^2\omega_2^2-1)\right)^2}{(\omega_1-\omega_2)^2(b^2\omega_1\omega_2+1)^2} \right).
\end{equation}
At the critical point this leads to the Casimir interaction free energy 
\begin{eqnarray}
f(h) = && \frac{k_BT}{2\pi h^2} \left(\frac{1}{2}\int_{0}^{\infty} \!\!\!\!\! dt\ t\ln{\left(1- \exp(-2t) \left( 1 - t\frac{b^2t^2-h^2}{b^2t^2+h^2}\right)^2\right)} +   \int_{0}^{h q_0} \!\!\!\!\! dt\ t\ln{\vert\sin{t}\vert}\right). 
\label{bakxruyb}
\end{eqnarray}

We see that in general the contributions from the modes $q>q_0$ depends on the precise value of $b$, that introduces an additional length scale, and consequently the simple $1/h^2$ part of the Casimir interaction is modified. In fact, the first integral in Eq. (\ref{bakxruyb}) depends only on the ratio $x = h/b$ 
\begin{equation}
 g(x) = \frac{1}{2}\int_{0}^{\infty} \!\!\!\!\! dt\ t\ln\left(1- \exp(-2t) \left( 1 - t\frac{t^2-x^2}{t^2+x^2}\right)^2\right)
\end{equation}
and its behavior is shown in Fig. \ref{fig:fig6p1}.

However, the contribution from the modes $q<q_0$ is independent of $b$ and is exactly the same as that derived for the confined field, in agreement with the observation made in Section \ref{basic} that at the critical point this must be the case in general due to Eq. (\ref{hdiff}). For $b=0$, that is to say in the case of Dirichlet-Dirichlet (D-D) boundary conditions, we find 
\begin{eqnarray}
f(h) &=& \frac{k_BT }{2\pi h^2} \left(\frac{1}{2}\int_{0}^{\infty} \!\!\!\!\! dt\ t\ln{\left(1-\exp(-2t)(1+t)^2\right)} +  \int_{0}^{h q_0} \!\!\!\!\! dt\ t\ln{\vert\sin{t}\vert}\right) \nonumber\\
&=& \frac{k_BT }{2\pi h^2} \left({-0.81726}+   \int_{0}^{h q_0} \!\!\!\!\! dt\ t\ln{\vert\sin{t}\vert}\right). 
\label{mwegh}
\end{eqnarray}
while for Neumann-Neumann (N-N) boundary conditions we have 
\begin{eqnarray}
f(h) &=&  \frac{k_BT }{2\pi h^2} \left(\frac{1}{2}\int_{0}^{\infty} \!\!\!\!\! dt\ t\ln{\left(1-\exp(-2t)(1-t)^2\right)} +  \int_{0}^{h q_0} \!\!\!\!\! dt\ t\ln{\vert\sin{t}\vert}\right) \nonumber\\
&=& \frac{k_BT }{2\pi h^2} \left( -0.0680951+   \int_{0}^{h q_0} \!\!\!\!\! dt\ t\ln{\vert\sin{t}\vert}\right). 
\label{bfgjq}
\end{eqnarray}
We thus see that the presence of external bulk has a strong influence on the amplitude of the Casimir interaction free energy generated by modes with $q>q_0$, but as predicted earlier has no influence on the modes with $q<q_0$. {The salient features of the Casimir interaction of the type Eqs. (\ref{mwegh}) and (\ref{bfgjq}) have been analysed before, see the discussion of Eqs. (\ref{befjkw0}) and (\ref{dnc}), and will not be repeated here.}

Dirichlet-Neumann (D-N) boundary conditions are again obtained via the limit {$b_1=0$, $b_2\to\infty$ }
which yields 
\begin{equation}
f(h) = \frac{k_B T }{2(2\pi)^2}\int d{\bf q}\ln{\left( 1 - \frac{\omega_1 \omega_2\left(\exp(-\omega_1h) -\exp(-\omega_2 h) \right)^2}{(\omega_1-\omega_2)^2} \right)},
\end{equation}
which at the critical point yields 
\begin{eqnarray}
f(h) &=& \frac{k_BT }{2\pi h^2} \left(\frac{1}{2}\int_{0}^{\infty} dt\ t\ln{\Big( 1- t^2 \exp(-2t)\Big)} + \int_0^{ hq_0}\!\!\!\!\!dt\  t\ \ln|\cos(t)|\right) \nonumber \\
&=& \frac{k_BT }{2\pi h^2} \left(-0.195371 + \int_0^{ hq_0}\!\!\!\!\!dt\  t\ \ln|\cos(t)| \right).
\end{eqnarray}
Again we see exactly the same contribution from the modes $q<q_0$ as the case of Dirichlet-Neumann boundary conditions for confined systems in Eq. (\ref{dnc}). However the contribution from modes with $q>q_0$ leads to an attractive $1/h^2$ interaction as opposed to the repulsive form seen in Eq. (\ref{dnc}).

\subsection{Some more exotic boundary conditions}

Here the  case where there is no boundary condition on any of the surfaces leads to no interaction. This is obviously correct, as in a nonconfined system removing the boundary conditions at a surface effectively removes that surface.

In the case of strong anchoring on one surface and Robin on the other, we find
\begin{eqnarray}
Z_U({\bf q},h) &=& \frac{(\omega_1\omega_2)(\omega_1+\omega_2)[(\omega_1^2-\omega_2^2)^2]^\frac{1}{2}\exp(\frac{1}{2} (\omega_1+\omega_2) h)}{2\pi^2 M^{\frac{1}{2}}} \int d\dot\phi\exp\left(-\frac{\dot\phi^2}{2}{\bf u}\cdot  B_1P S_{DR}P B_1 {\bf u}\right)\nonumber \\
&=& \frac{(\omega_1\omega_2)(\omega_1+\omega_2)[(\omega_1^2-\omega_2^2)^2]^\frac{1}{2}\exp(\frac{1}{2} (\omega_1+\omega_2) h)}{\pi(2\pi)^{\frac{1}{2}} M^{\frac{1}{2}} ({\bf u}\cdot B_1 P S_{DR}P B_1{\bf u})^{\frac{1}{2}} }.
\end{eqnarray}
In the limit of Dirichlet boundary conditions on the Robin surface (DN-D), and at the critical point we find
\begin{eqnarray}
f(h) &=& \frac{k_BT }{2\pi h^2} \left(\frac{1}{2}\int_{0}^{\infty} dt\ t\ln{\Big( 1- \exp(-2t)(1+2t+2t^2)\Big)} + \frac{1}{2}\int_0^{q_0h}dt\  t \ln\left(2t-\sin(2t)\right)\right] \nonumber \\
&=& \frac{k_BT }{2\pi h^2} \left({-1.20552} +\frac{1}{2}\int_0^{q_0h}dt\  t \ln\left(2t-\sin(2t)\right) \right).
\label{efjkw}
\end{eqnarray}
In the limit of Neumman boundary conditions on the Robin surface (DN-N), and at the critical point, one finds
\begin{eqnarray}
f(h) &=& \frac{k_BT }{2\pi h^2} \left(\frac{1}{2}\int_{0}^{\infty} dt\ t \ln{\Big( 1- \exp(-2t)(1-2t+2t^2)\Big)}+ \frac{1}{2}\int_0^{q_0h}dt\  t \ln\left(2t+\sin(2t)\right)\right] \nonumber \\
&=& \frac{k_BT }{2\pi h^2} \left(-0.266976 + \frac{1}{2}\int_0^{q_0h}dt\  t \ln\left(2t+\sin(2t)\right) \right).
\label{pijfa}
\end{eqnarray}
Again, the Casimir interaction of the type Eqs. (\ref{efjkw}) and (\ref{pijfa}) has been discussed in Section \ref{Secexpot}.

\section{Discussion}

We have considered a second order derivative field theory of the general Brazovskii type, Eq. (\ref{form1}), in the presence of parallel plates, which modify the field fluctuations. The field theory has a critical point at $p_0=0$ and the bulk has a continuum of  zero modes. We thus expect the presence of a critical thermal long range Casimir interaction at the critical point. Two distinct cases have been considered, the confined field case and the unconfined field case. These two cases differ by the presence of surface terms, which can strongly modify the form of the corresponding Casimir free energy, from which one derives the Casimir component of the disjoining pressure.

In all cases, at the critical point, we find that the Casimir free energy per unit area can be written as
\begin{equation}
f(q_0 h)= \frac{q_0^2k_B T}{2\pi (q_0 h)^2}\left[ H + \frac{1}{2}\int_0^{q_0 h} dt ~t~ \ln(r(t))\right].
\label{cagr}
\end{equation}
Here $H$ is an effective amplitude or Hamaker coefficient \cite{Parsegian}, which depends on the boundary conditions and whether the system is confined or not. The modes generating this term are the modes with $q>q_0$. The second term is the one that can introduce oscillatory behavior. The function $r(t)$ can take several forms with $r(t)>0$, but it can have an infinite number of zeros leading to an infinite number of metastable states. Remarkably, in contrast to what is found for the amplitude $H$, the form of $r(t)$ is independent of confinement. 
\begin{table}[t!]
  \begin{tabular}{  | c | c |c | c| }
    \hline
     & $H$ confined  & $H$ unconfined & $r(t)$ confined and unconfined \\ \hline \hline
    DN-DN & -1.71629 & -1.71629 & $t^2-\sin^2(t)$ \\ \hline
    D-D & -0.30051 & {-0.81726} &  $\sin^2(t)$ \\\hline
    N-N & -0.30051 &{-0.068095} &  $\sin^2(t)$ \\ \hline
    D-N & 0.225386 & {-0.195371} & $\cos^2(t)$\\ \hline
    DN-D & -0.83591& {-1.20552}& $2t-\sin(2t)$ \\ \hline
    DN-N & 0.406839 &-0.266976 & $2t+\sin(2t)$ \\
    \hline
    \end{tabular}
    \caption{Summary of the components of the Casimir free energy, Eq. \ref{cagr}, for various boundary conditions comparing confined and unconfined systems.}\label{tab}
    \end{table}

To summarize our results, the value of $H$ and the form of $r(t)$ are given Table \ref{tab}, from where one can compare the effect of confinement in various cases. Note that the cases of periodic/antiperiodic boundary conditions are omitted as they are not natural in the context of unconfined systems.

The above summary highlights two important points
\begin{itemize}
\item The coefficient $H$ which determines the Casimir  potential for small values of $h$, or equivalently small $q_0$, is always attractive for unconfined fields. For confined fields, the $q>q_0$ components are always attractive for symmetric boundary conditions. However, the non symmetric cases of DN-D and DN-N are attractive and repulsive, respectively, and D-N is repulsive.
\item In systems with a single boundary condition on each surface, the component coming from modes $q<q_0$ give an infinite number (periodic) of points, where the disjoining pressure diverges, this leads to an infinite number of metastable states in the system. No such divergence occurs when three or more boundary conditions are imposed (DN-D, DN-N and DN-DN).
\end{itemize}

{The behavior of the critical Casimir interaction for the general Brazovskii-type confined field theories with an additional length scale given by $q_0$ is thus quite distinct from the critical Casimir interactions (CCI) \cite{Fisher} studied in great detail in the context of collective behavior of colloids \cite{Gambassi, DietricCasimir}. In fact the critical Casimir interactions in the Brazovskii-type field theories can be seen as a generalization of the CCI applicable to the case of semi-flexible polymers \cite{Bruinsma2013, pap, doi, dob2001, smith, uchida}, soft membranes  \cite{boal, dean2007, Bing2017}, ionic liquids \cite{sant2006, ciach2007, bazant2011, blossey2017} and liquid crystals \cite{had_2018, had_2019}, where higher derivative Hamiltonians emerge naturally. The main difference between the CCI and the general Brazovskii-type confined field theories boils down to the fact that the  modes parallel to the bounding surfaces contribute to the Casimir interaction disjoining pressure in a manner  quite different to that in the CCI. The modes with $q<q_0$ can lead to oscillatory terms in the pressure, and exhibit longer range interactions than the CCI, giving a contribution to the disjoining pressure which decays with an inverse first power of the separation between the bounding surfaces, as opposed to the usual inverse third power characteristic of the CCI as well as the Lifshitz - van der Waals interactions \cite{Parsegian}. This long range effect and the non-monotonic dependence of the disjoining pressure on the separation between bounding surfaces, creating a possibility of multiple metastable states, makes the fundamental study of critical Casimir interaction for the general Brazovskii-type confined field theories quite interesting.}

\section{Acknowledgment}

DSD acknowledges support from the ANR Grants FISICS and RaMaTraF. BM acknowledges support from the National Natural Science Foundation of China (NSFC) (Grant Nos. 21774131 and 21544007). RP acknowledges support of the {\sl 1000-Talents Program} of the Chinese Foreign Experts Bureau, and of the University of the Chinese Academy of Sciences, Beijing.


%

\end{document}